\documentclass[3p,times,twocolumn]{elsarticle}

\usepackage{ecrc}

\volume{00}

\firstpage{1}

\journalname{Nuclear and Particle Physics Proceedings}

\runauth{}

\jid{nppp}

\jnltitlelogo{Nuclear and Particle Physics Proceedings}

\usepackage{amssymb}

\usepackage[figuresright]{rotating}

\usepackage{colordvi}
\usepackage[normalem]{ulem}

\usepackage{color}
\newcommand{\beqn}{\begin{eqnarray}}
\newcommand{\eeqn}{\end{eqnarray}}
\newcommand{\be}{\begin{equation}}
\newcommand{\ee}{\end{equation}}

\newcommand{\ba}{\begin{array}{c}}
\newcommand{\bat}{\begin{array}{cc}}
\newcommand{\ea}{\end{array}}

\newcommand{\bi}{\begin{itemize}}
\newcommand{\ei}{\end{itemize}}

\newcommand{\ket}{\,\rangle}
\newcommand{\bra}{\langle \,}

\newcommand{\Frac}[2]{\frac{\displaystyle #1}{\displaystyle #2}}

\newcommand{\cO}{{\cal O}}

\newcommand{\mF}{\mathcal{F}}

\newcommand{\mL}{\mathcal{L}}

\newcommand{\mO}{\mathcal{O}}

\newcommand{\mY}{\mathcal{Y}}

\newcommand{\gsim}{\stackrel{>}{_\sim}}

\newcommand{\comment}[1]{}

\begin{document}

\begin{frontmatter}



\dochead{}

\title{Electroweak effective theory and beyond Standard Model resonances}


\author{Juan Jos\'e Sanz-Cillero\corref{cor0} }
\ead{jjsanzcillero@ucm.es} 
\address{Departamento de F\'\i sica Te\'orica and IPARCOS, Universidad Complutense de Madrid, E-28040 Madrid, Spain}

\author{Antonio Pich}
\ead{pich@ific.uv.es} 
\address{IFIC, Universitat de Val\`encia – CSIC, Apt. Correus 22085, 46071 Val\`encia, Spain}

\author{Ignasi Rosell}
\ead{rosell@uchceu.es} 
\address{Departamento de Matemáticas, F\'\i sica y Ciencias Tecnol\'ogicas, Universidad Cardenal Herrera-CEU, CEU Universities, 46115 Alfara del Patriarca, València, Spain}

\cortext[cor0]{Speaker. Talk given at {\it the 23rd International Conference in Quantum Chromodynamics: QCD20 - 35 years later} (QCD 20), 27-30 October 2020, Montpellier, France. 
We wish to thank the organizers of the  conference. 
Work supported by the Spanish
Government and ERDF funds from the European Commission (FPA2016-75654-C2-1-P, FPA2017-84445-P, PID2019-108655GB-I00),   
Generalitat Valenciana (PROMETEO/2017/053),  
Universidad Cardenal Herrera-CEU
(INDI18/11, INDI19/15),  
the EU STRONG-2020 project  
[grant no. 824093], and the STSM Grant from COST Action CA16108.}

\begin{abstract} 
We consider a non-linear realization of the electroweak symmetry-breaking pattern $SU(2)_L\times SU(2)_R/SU(2)_{L+R}$ to construct a low-energy effective theory, later extended by the inclusion of heavy new-physics resonances. After assuming appropriate high-energy constraints given by Weinberg sum-rules and the asymptotic behaviour of form-factors, we obtain relations between resonance masses and some low-energy effective couplings. These predictions are compared with current experimental data and some resonance mass bounds are inferred. 
\end{abstract}


\end{frontmatter}
 
\section{Electroweak effective theory and resonance extensions}

In these proceedings we discuss the effective field theory (EFT) description of the electroweak (EW) gauge and Higgs sectors and its phenomenological application put forward  
in Ref.~\cite{Pich:2020xzo}. We adopt  the non-linear realization of the EW symmetry provided by the electroweak effective theory (EWET)~\cite{Krause:2018cwe},  
also known as Higgs effective field theory (HEFT) or electroweak chiral Lagrangian (EWChL)~\cite{deFlorian:2016spz,Pich:2018ltt,EFT,Longhitano:1980iz,Herrero:1993nc}.   
The basic ingredient of the EFT are the SM symmetries, in particular the EW symmetry-breaking pattern $SU(2)_L\times SU(2)_R/SU(2)_{L+R}$, where the chiral group gets spontaneously broken down to the custodial subgroup.\footnote{This exact symmetry of the scalar sector is softly broken in the full Standard Model when the fermion and EW gauge-boson interactions are plugged in.}

The effective Lagrangian is organized in growing powers of $p$, standing $p$ for any soft scale of the EFT (external momenta, masses $m_h,\, m_W, $ etc.):
\begin{equation}
\mL_{\rm EWET} \,=\, \sum_{\hat{d}\geq 2} \mL^{(\hat{d})}_{\rm EWET}\, .
\end{equation}

The leading order (LO) term is of $\mO(p^2)$. In addition to all the Standard Model (SM) operators, $\mL^{(2)}_{\rm EWET}$ contains some new physics interactions~\cite{Pich:2018ltt,EFT,Longhitano:1980iz,Herrero:1993nc}, 
\begin{eqnarray} 
    \label{eq:L.LO}
&&\hspace*{-1.35cm}
\mathcal{L}_{\mathrm{EWET}}^{(2)} \, =\, \sum_{\xi} \left[ i\,\bar\xi \gamma^\mu d_\mu \xi -v \left( \; \bar\xi_L\, \mY\, \xi_R \, +\, \mathrm{h.c.} \right) \right] 
 \\
&&\hspace*{-1.35cm}
 -\frac{1}{2g^{2}}\langle \hat{W}_{\mu\nu}\hat{W}^{\mu\nu}\rangle_{2}-\frac{1}{2g^{'2}}\langle \hat{B}_{\mu\nu}\hat{B}^{\mu\nu}\rangle_{2}-\frac{1}{2g_{s}^{2}}\langle \hat{G}_{\mu\nu}\hat{G}^{\mu\nu}\rangle_{3} 
\nonumber\\
&&\hspace*{-1.35cm}
 +\frac{1}{2}\partial_\mu h \partial^\mu h  -\frac{1}{2} m_h^2 h^2  - V(h/v)  + \frac{v^2}{4}\mF_u(h/v)\,\langle u_\mu u^\mu\rangle_{2}\, , 
\nonumber
\end{eqnarray}
where the EW Goldstones $\varphi^a$ are non-linearly realized through the usual exponential parametrization $u(\varphi) = \exp(i\vec{\sigma}\vec{\varphi}/(2v))$,  
with the Pauli matrices $\sigma^a$ and the EW scale $v = \left(\sqrt{2} G_F\right)^{-1/2} = 246\:\mathrm{GeV}$.  
The Higgs field $h$ is a singlet and can enter through undetermined functions $\mF(h/v)$ that may appear in front of the chiral operators. In particular,  
the Goldstone kinetic term introduces the factor 
\begin{equation}
\mF_u(h/v)\,=\, 1+ \frac{2 \kappa_W h}{v} + \frac{c_{2V} h^2}{v^2} + \cO(h^3)\, .
\end{equation} 
The couplings $\kappa_W$ and $c_{2V}$ parametrize the $hWW$ and $hhWW$
interactions. These low-energy constants (LEC) are normalized such that  
$\kappa_W=c_{2V}=1$ in the SM. An alternative notation 
$a = \kappa_W$ and $ b = c_{2V}$ is used in some works.

At next-to-leading order (NLO), $\cO(p^4)$, the effective Lagrangian has the structure~\cite{Krause:2018cwe},
\begin{eqnarray}
  \label{eq:L-NLO}
\mathcal{L}_{\mathrm{EWET}}^{(4)} & =&
\sum_{i=1}^{12} \mF_i(h/v)\; \mO_i \, +\, \sum_{i=1}^{3}\widetilde\mF_i(h/v)\; \widetilde \mO_i  \nonumber
\nonumber\\ &&
\hspace*{-1.5cm}\, +\,  \sum_{i=1}^{  8  } \mF_i^{\psi^2}(h/v)\; \mO_i^{\psi^2} \, +\, \sum_{i=1}^{  3   } \widetilde\mF_i^{\psi^2}(h/v)\; \widetilde \mO_i^{\psi^2}
\nonumber \\ &&
\hspace*{-1.5cm}\, +\, \sum_{i=1}^{10}\mF_i^{\psi^4}(h/v)\; \mO_i^{\psi^4} \, +\, \sum_{i=1}^{2}\widetilde\mF_i^{\psi^4}(h/v)\; \widetilde \mO_i^{\psi^4} \, .
\end{eqnarray} 
The precise form of these NLO operators can be found in Refs.~\cite{Pich:2020xzo,Pich:2018ltt}.
Any non-zero value of the $\cO(p^4)$ LECs would signal the presence of new physics (NP).
Some examples of NLO computations within the EWET can be found in Refs.~\cite{deFlorian:2016spz,NLO-computations,Fabbrichesi:2015hsa}.

Only the $CP$--even bosonic sector is studied in this work~\cite{Pich:2020xzo}. 
Thus, we will focus our attention on  the purely bosonic operators of the EFT Lagrangian, the structures $\mF_i(h/v) \, \mO_i$ and $\widetilde{\mF}_i(h/v) \, \widetilde{\mO}_i$ . For $h = 0$, the $\mF_i(0)$ and $\widetilde{\mF}_i(0)$ LECs are related to the $a_i$ couplings of the Higgsless Longhitano Lagrangian~\cite{Longhitano:1980iz,Herrero:1993nc} in the form: $a_i=\mF_i$ for $i=1,4,5$, $a_2=(\mF_3-\widetilde{\mF}_1)/2$ and $a_3=-(\mF_3+\widetilde{\mF}_1)/2$.

We extend the validity of our EFT by including the lightest NP degrees of freedom, heavy resonances with masses $M_R\gg v$. Since we are interested in their impact on the purely bosonic part of the low-energy effective Lagrangian up to $\cO(p^4)$, only $J^{PC}$ resonances with spin $J=0$ and $J=1$  are discussed here: $0^{++}\, (S)$, $0^{-+}\, (P)$, $1^{--}\, (V)$ and $1^{++}\, (A)$. 
Other more exotic resonances that might contribute, such as tensor resonances, are discussed in Refs.~\cite{Donoghue:1988ed},   
finding analogous conclusions to those presented in Ref.~\cite{Pich:2020xzo}.  
Based on a generating functional approach~\cite{Ecker:1988te},  
the heavy resonance fields are 
integrated out from the action to extract the tree-level contributions to the low-energy EFT. Up to NLO in the EWET, $\cO(p^4)$, they are 
determined by  
terms containing one resonance field and an $\cO(p^2)$ chiral tensor: 
\begin{eqnarray}
\lefteqn{\Delta \mathcal{L}_{\mathrm{RT}} =\,
 \frac{v^2}{4}\!\left(\! 1 \!+\!\Frac{2\,\kappa_W}{v} h \!+\! c_{2V} \,h^2\!\right)\! \bra u_\mu u^\mu\ket_2} && 
\nonumber\\
&&+\, \Frac{c_{d}}{\sqrt{2}}\, S^1_1\bra u_\mu u^\mu \ket_2  + d_P \Frac{(\partial_\mu h)}{v} \bra  P^1_3  \, u^\mu \ket_2
\nonumber\\ 
&& 
+\,\widetilde{c}_{\mathcal{T}}\,  \hat V^1_{1\,\mu}  \bra u^\mu \mathcal{T} \ket_2  
+ c_{\mathcal{T}} \, \hat A^1_{1\,\mu}  \bra u^\mu \mathcal{T} \ket_2  \phantom{\bigg( }
 \nonumber \\
&&+\,\bra V^1_{3\,\mu\nu} \bigg( \Frac{F_V}{2\sqrt{2}}  f_+^{\mu\nu} + \Frac{i G_V}{2\sqrt{2}} [u^\mu, u^\nu]  \nonumber\\
&&
\qquad\qquad +\, \Frac{\widetilde{F}_V }{2\sqrt{2}} f_-^{\mu\nu}  +  \sqrt{2}\, \widetilde{\lambda}_1^{hV}   (\partial^\mu h) u^\nu  \bigg) \ket_2  
\nonumber\\
&& +\,   F_{X} V^1_{1\,\,\mu\nu}  \hat X^{\mu\nu} + C_G V^8_{1\,\mu\nu}  \hat G^{\mu\nu}\nonumber \\
&&+\, \bra A^1_{3\,\mu\nu} \bigg(\Frac{F_A}{2\sqrt{2}}  f_-^{\mu\nu}  + \sqrt{2}\, \lambda_1^{hA}  (\partial^\mu h) u^\nu 
\nonumber\\
&&
\qquad\qquad +\,  \Frac{\widetilde{F}_A}{2\sqrt{2}} f_+^{\mu\nu} +  \Frac{i \widetilde{G}_A}{2\sqrt{2}} [u^{\mu}, u^{\nu}]    \bigg) \ket_2  
\nonumber\\
&& +\,  \widetilde{F}_{X} A^1_{1\,\mu\nu}  \hat X^{\mu\nu} + \widetilde C_G  A^8_{1\,\mu\nu} \hat G^{\mu\nu} \,.
 \label{Lagrangian}
\end{eqnarray}
Only the terms which contribute to the $\cO(p^4)$ bosonic LECs are displayed here. 
The spin--1 resonances can be described with either a four-vector Proca field $\hat{R}^\mu$ or with an antisymmetric tensor $R^{\mu \nu}$~\cite{Ecker:1988te}.  
Here we keep both formalisms because, as it was demonstrated in Ref.~\cite{Krause:2018cwe}, the sum of tree-level resonance-exchange contributions from the $\cO(p^2)$ resonance Lagrangian with Proca and antisymmetric spin-1 resonances gives the complete set of predictions for the $\cO(p^4)$ EWET LECs, without any additional contributions from local $\cO(p^4)$ operators in the high-energy theory without resonance fields.

The detailed form of the contributions to the bosonic $\cO(p^4)$ LECs can be found in Refs.~\cite{Krause:2018cwe}.

\section{Ultraviolet completion}

Considered without any further information, our extended resonance theory is not really of much use yet, as we have simply traded various unknown LECs by several unknown resonance parameters. 
In order to become predictive one needs to make further assumptions on the behaviour of the underlying theory in the ultraviolet (UV). In our work~\cite{Pich:2020xzo}, we  
assume that the vector and axial-vector form-factors into two scalars ($\varphi\varphi$ and $h\varphi$) 
vanish at high energies, yielding the constraints
\begin{eqnarray}
v^2\, -\,F_V\,G_V\,   -\,\widetilde{F}_A\,\widetilde{G}_A \,=\, 0 \,. \label{VFF1}
\\
\widetilde{F}_V\, G_V  \, +\, F_A\,\widetilde{G}_A \,=\, 0 \,. \label{AFF1} 
\\
\kappa_W\,v \, -\,F_A\,\lambda_1^{hA}\,   -\,\widetilde{F}_V\,\widetilde{\lambda}_1^{hV} \,=\, 0\,, \label{AFF2}
\\
\widetilde{F}_A\, \lambda_1^{hA} \, +\, F_V\,\widetilde{\lambda}_1^{hV} \,=\, 0 \,. \label{VFF2}
\end{eqnarray}

In addition we  
study the 1st and 2nd Weinberg sum-rules~(WSR)~\cite{Weinberg:1967kj} 
for the $W^3B$ correlator~\cite{Peskin:1990zt},  
obtaining, respectively, the high-energy relations 
\begin{eqnarray}
F_V^2 +\widetilde{F}_A^2 - F_A^2 - \widetilde{F}_V^2 \,=\, v^2  \,. \label{1WSR}
\\
 F_V^2 M_V^2 + \widetilde{F}_A^2 M_A^2 - F_A^2 M_A^2 - \widetilde{F}_V^2 M_V^2\, =\, 0\, , \label{2WSR}
\end{eqnarray}
which imply~\cite{Pich:2020xzo}  
\begin{eqnarray}
\label{eq:FVFA-WSRs}
&&\hspace*{-1.4cm}
F_V^2 - \widetilde{F}_V^2\, =\, \frac{v^2 M_A^2}{M_A^2-M_V^2}\, ,
\quad
F_A^2 - \widetilde{F}_A^2 \,=\, \frac{v^2 M_V^2}{M_A^2-M_V^2}\, , 
\end{eqnarray}
with $M_A>M_V$. Nevertheless, in scenarios such as Conformal and Walking Technicolour models, the 2nd WSR cannot be applied~\cite{Orgogozo:2011kq}  
and the relations~(\ref{eq:FVFA-WSRs}) are no-longer applicable. 
Thus, we will also study the implications derived from assuming only the 1st WST, Eq.~(\ref{1WSR}).

\begin{table}[!t]  
\begin{center}
\renewcommand{\arraystretch}{2.6}
\begin{tabular}{|c|c|c|}
\hline
& \multicolumn{2}{|c|}{$\mF_i$} \\ \hline
$i$ & with 2nd WSR & without 2nd WSR
\\ \hline\hline
$1$ &  {\small   $-\displaystyle\frac{v^2}{4} \left( \frac{1}{M_V^2} \!+\! \frac{1}{M_A^2} \right)$ }   
&   {\small    $\!-\displaystyle\frac{v^2}{4M_V^2}\! -\!\frac{F_A^2}{4} \!\left(\!\frac{1}{M_V^2}\!-\!\frac{1}{M_A^2}\!\right) < 
\displaystyle\frac{-v^2}{4M_V^2}$   }  \\[1ex] \hline%
$3$ & \multicolumn{2}{|c|}{$-\displaystyle\frac{v^2}{2M_V^2}$} 
\\[1ex] \hline
$4$ & $\displaystyle\frac{v^2}{4} \left( \frac{1}{M_V^2}\!-\!\frac{1}{M_A^2}\right)$& $\cdots$ 
\\[1ex] \hline 
$5$ & \multicolumn{2}{|c|}{$\displaystyle\frac{c_d^2}{4M_{S^1_1}^2}-\mathcal{F}_4$}
\\[1ex] \hline
$6$ &  $-\displaystyle \kappa_W^2 v^2\left(\frac{1}{M_V^2}\! -\! \frac{1}{M_A^2} \right)$& 
$\cdots$ 
\\[1ex] \hline
$7$ &  \multicolumn{2}{|c|}{$\displaystyle\frac{d_P^2}{2M_P^2} -\mathcal{F}_6$}
\\[1ex] \hline
 $9$  &   \multicolumn{2}{|c|}{$-\displaystyle\frac{\kappa_W v^2}{M_A^2} $}
\\[1ex] \hline
\end{tabular}
\caption{{\small
 Resonance-exchange contributions to the $P$-even bosonic $\cO(p^4)$ LECs, considering only $P$-even operators and the short distance constraints. 
 }}
\label{p_even_UV}
\end{center}
\end{table}

\begin{table}[!t]  
\begin{center}
\renewcommand{\arraystretch}{2.6}
\begin{tabular}{|c|c|c|}    
\hline
& \multicolumn{2}{|c|}{$\mF_i$} \\ \hline
{\small $i$} & {\small with 2nd WSR} & {\small without 2nd WSR}
\\ \hline\hline
1 &  {\small  $\quad  -\displaystyle\frac{v^2}{4} \left( \frac{1}{M_V^2} + \frac{1}{M_A^2} \right)$  }   & {\small $ -\displaystyle\frac{v^2}{4M_V^2} -\frac{F_A^2-\widetilde F_A^2}{4}  \left( \frac{1}{M_V^2}\! -\!\frac{1}{M_A^2} \right)$  }
\\ &&
{\small ${}^\dagger <  -\displaystyle\frac{v^2}{4M_V^2}$ } \\[1ex] \hline
3 &    \multicolumn{2}{|c|}{ {\small   $-\displaystyle\frac{v^2}{2M_A^2} - \frac{F_VG_V}{2} \left( \frac{1}{M_V^2} \!-\! \frac{1}{M_A^2} \right)
\quad {}^\ddagger < \quad -\displaystyle\frac{v^2}{2M_A^2}
$ }   }     
\\[1ex] \hline
5 &  \multicolumn{2}{|c|}{  {\small   $ \displaystyle\frac{c_d^2}{4M_{S^1_1}^2} -\mathcal{F}_4 \phantom{\Bigg)}$   }  }
\\[1ex] \hline
7 &  \multicolumn{2}{|c|}{ {\small  $ \displaystyle\frac{d_P^2}{2M_P^2} -\mathcal{F}_6$  } }
\\[1ex] \hline
9 &    \multicolumn{2}{|c|}{   {\small  $-\displaystyle\frac{\kappa_W v^2}{M_V^2} + F_A \lambda_1^{hA}v \left(  \frac{1}{M_V^2}\!-\!\frac{1}{M_A^2} \right)
\quad {}^\mathsection > \quad -\displaystyle\frac{\kappa_W v^2}{M_V^2}$  }   }
\\[1ex] \hline
\end{tabular}
\caption{{\small
Resonance-exchange contributions to the $P$-even bosonic $\cO(p^4)$ LECs ($P$-even and $P$-odd operators included), once the short distance constraints are considered. 
The inequalities $^\dagger$, $^\ddagger$ and $^\mathsection$ assume 
that $F_A^2>\widetilde F_A^2$, $F_VG_V>0$ and $F_A \lambda_1^{hA}>0$, respectively.}
}
\label{p_even_odd_UV}
\end{center}
\end{table}

\section{Low-energy constant predictions}

In this Section we apply our 
UV assumptions in the previously discussed LEC predictions~\cite{Krause:2018cwe}.   
The form-factor and WSR constraints~(\ref{VFF1})--(\ref{eq:FVFA-WSRs}) imply relations between resonance couplings in such a way that some of the LECs  
can be now expressed in terms of the resonance masses $M_R$ or can be bounded by specific combinations of them. 
As previously pointed out, we  
also perform a separate analysis for the case when the 2nd WSR is discarded, which  
obviously makes our results less predictive.

The implications of the UV constraints on the bosonic $\cO(p^4)$ LECs are summarized in Tables~\ref{p_even_UV} and~\ref{p_even_odd_UV}, 
with and without assuming the 2nd WSR. 
In addition, we first study the case when the underlying high-energy theory preserves parity,  Table~\ref{p_even_UV}, and then we discuss in Table~\ref{p_even_odd_UV} the impact of allowing parity violation. In some cases, it is not possible to extract any prediction, which 
we indicate with the symbol ``$...$''. In other cases, only bounds can be extracted. In order to obtain these inequalities in the scenario with $P$--odd operators we needed to assume that parity breaking, though not negligible, was nevertheless suppressed. Further details can be found in Ref.~\cite{Pich:2020xzo}. 

In order to ease the notation, and taking into account that most contributions come from resonances in EW triplets and  QCD singlets ($R^1_3$), neither superindices nor subindices are used in this case in Tables~\ref{p_even_UV} and~\ref{p_even_odd_UV}, and from now on; that is, $M_R\equiv M_{R^1_3}$ here.

\begin{table}[!t]  
\begin{center}
\renewcommand{\arraystretch}{1.2}
\begin{tabular}{|r@{$\,<\,$}c@{$\,<\,$}l|c|c| }
\hline
\multicolumn{3}{|c|}{LEC} & Ref. & Data \\ 
\hline \hline
$0.89$ & $\kappa_W$ & $1.13$  & \cite{deBlas:2018tjm}
& LHC  \\ \hline 
$-1.02$ & $c_{2V}$ & $2.71$ & \cite{ATLAS:2019dgh} & LHC \\ \hline
 $-0.004$  &$ \mathcal{F}_1$& $0.004$  & 
 \cite{Tanabashi:2018oca} & LEP via $S$\\ \hline 
  $-0.06$&$\mathcal{F}_3$ &$0.20$&\cite{Almeida:2018cld} & LEP \& LHC  \\ \hline 
 $-0.0006$&$\mathcal{F}_4$&$0.0006$&\cite{Sirunyan:2019der} & LHC  \\ \hline
$-0.0010$&$\mathcal{F}_4+\mathcal{F}_5$&$0.0010$ &  \cite{Sirunyan:2019der} & LHC  \\  \hline

\end{tabular}
\caption{{\small
Current experimental constraints on bosonic EWET LECs, at 95\% CL.}} \label{exp}
\end{center}
\end{table}

\section{Implications: $M_R$ bounds} 

Now we compare the current experimental knowledge on some of these couplings with the previous predictions and extract conclusions on the allowed values of the NP resonance masses. For this we took the 95\% confidence level (CL) inputs, given  in Table~\ref{exp}~\cite{Pich:2020xzo}:
\begin{enumerate}
\item $\kappa_W$ ($h\to WW$): CMS+ATLAS analysis within the HEFT framework~\cite{deBlas:2018tjm}.
\item $c_{2V}$ ($hh\to WW$): ATLAS bounds~\cite{ATLAS:2019dgh}. 
\item $\mathcal{F}_1$ ($W^3B$): $S$--parameter at LEP~\cite{Tanabashi:2018oca}.
\item $\mathcal{F}_3$ ($\gamma WW$): anomalous triple gauge coupling (TGC) $\delta \kappa_\gamma$ at LEP and LHC~\cite{Almeida:2018cld}. 
\item $\mathcal{F}_4$ and $\mathcal{F}_5$ ($WW\to WW$): CMS vector boson scattering (VBS)~\cite{Sirunyan:2019der}.\footnote{
A more recent CMS measurement~\cite{Sirunyan:2020gyx}  
finds indeed significantly less stringent constraints when the unitarity requirements are taken into consideration~\cite{Fabbrichesi:2015hsa,Garcia-Garcia:2019oig}. These caveats are discussed in Ref.~\cite{Pich:2020xzo}. 
}
\end{enumerate} 

In order to compare these results with our theoretical LEC determinations, 
together with the experimental confidence intervals in Table~\ref{exp},
one must also take into account the one-loop EWET uncertainties  
which were estimated to be~\cite{ Pich:2020xzo}: 
\begin{eqnarray}
&&\hspace*{-1.25cm} \Delta \mathcal{F}_1 = \Delta \mF_3 =  0.9 \cdot 10^{-3}\, ,
\quad\;\;\: 
\Delta \mathcal{F}_{4} =  3 \cdot 10^{-5}\, ,
\nonumber\\
&&\hspace*{-1.25cm} \Delta( \mathcal{F}_{4}+\mathcal{F}_{5}) = 1.7 \cdot  10^{-3}\, , 
\qquad\;\;\,    
\Delta \mathcal{F}_{6} =  3\cdot  10^{-3}\, ,
\nonumber\\
&&\hspace*{-1.25cm} \Delta (\mathcal{F}_{6}+\mF_7) =  0.6\cdot  10^{-2}\, , 
\qquad  
\Delta \mathcal{F}_{9} =  1.4\cdot  10^{-2}\, . \quad
\label{eq:1loop-error}
\end{eqnarray}
In these proceedings we will focus on the  
couplings $\mF_{1,3,4,5}$, whose NLO uncertainties above only depend on $\kappa_W$. In Ref.~\cite{Pich:2020xzo}, we have also studied the yet-to-be-measured couplings $\mF_{6,7,9}$ (related to Higgs production processes), which have much larger one-loop uncertainties as they also depend on the poorly known $hhWW$ coupling $c_{2V}$. All the estimates in~(\ref{eq:1loop-error}) vanish in the SM limit $\kappa_W=c_{2V}=1$.

\begin{figure*}[!t]
\begin{center}
\begin{minipage}[c]{6.4cm}
\includegraphics[width=6.4cm]{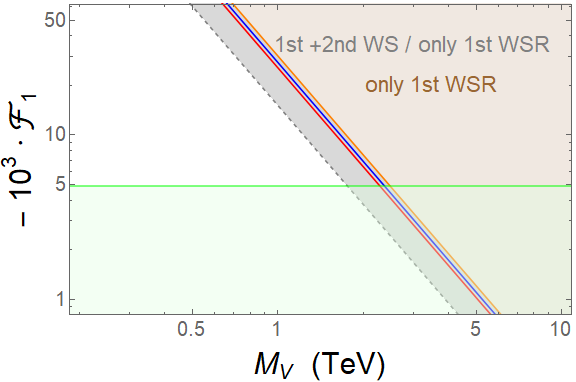}  
\end{minipage} 
\hspace*{1cm}
\begin{minipage}[c]{6.4cm}
\includegraphics[width=6.4cm]{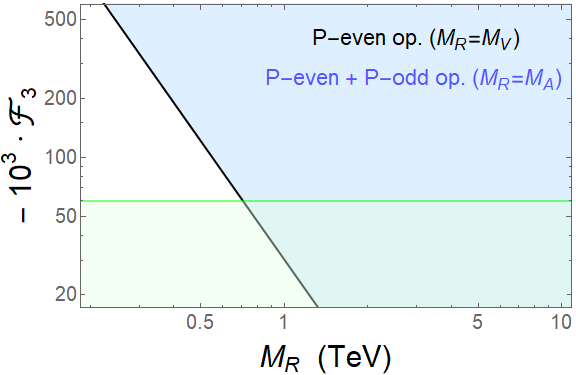} 
\end{minipage}
\\[8pt]
\begin{minipage}[c]{6.4cm}
\includegraphics[width=6.4cm]{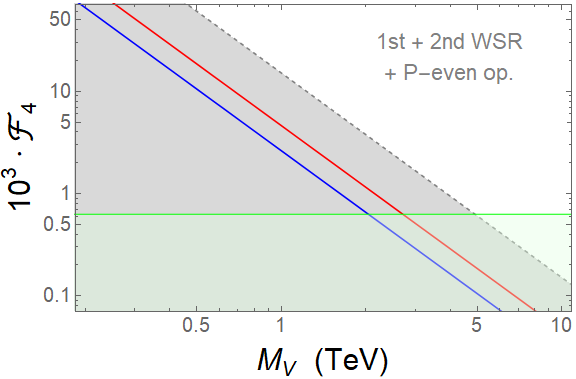} 
\end{minipage} 
\hspace*{1cm}
\begin{minipage}[c]{6.4cm}
\includegraphics[width=6.4cm]{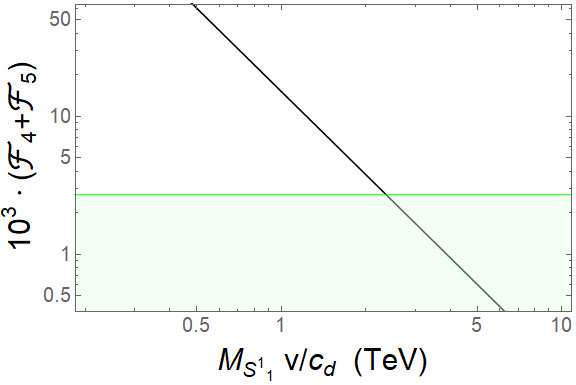}
\end{minipage}
\end{center}
\caption{{\small Predicted values for the LECs $\mF_1$, $\mF_3$, $\mF_4$ and $\mF_4+\mF_5$, 
from Tables~\ref{p_even_UV} and~\ref{p_even_odd_UV}, 
as a function of the corresponding resonance mass ($M_V$, $M_A$ or $M_{S_1^1} \,v/c_d$). The green area covers the combination of the 95\% CL experimentally allowed region and the estimated one-loop uncertainties in Eq.~(\ref{eq:1loop-error}). If there is a dependence on $M_V$ and $M_A$, the gray and/or brown regions cover all possible values for $M_A>M_V$. If the 2nd WSR has been considered, it is explicitly indicated in the plot, with the corresponding lines for $M_A=M_V$ (orange), $M_A=1.1\,M_V$ (blue), $M_A=1.2\,M_V$ (red) and $M_A\to \infty$ (dashed dark gray). 
In the case without the 2nd WSR, the theoretically allowed region for $\mF_1$ is given by both the gray and brown regions. The $\mF_3$ result does not rely on any WSR. In case of using only the even-parity operators, we indicate it in the plot.
}} 
\label{fig:plots1}
\end{figure*}

The implementation of short-distance constraints has allowed us to determine some bosonic LECs in terms of very few resonance parameters, as shown in Tables~\ref{p_even_UV} and~\ref{p_even_odd_UV}. 
The comparison of our theoretical predictions for the LECs $\mF_{1,3,4,5}$ and their experimentally-allowed values is provided in Fig.~\ref{fig:plots1}.


Observing the plots in Fig.~\ref{fig:plots1}, one can extract the following conclusions:
\begin{itemize}
    \item $S$--parameter data: $\mF_1$ implies $M_{V,A}\gsim 2$~TeV.
    \item Anomalous TGC: $\mF_3$ implies $M_{V,A}\gsim 0.5$~TeV.
    \item VBS: $\mF_4$ and $\mF_5$ at the precision level of Ref.~\cite{Sirunyan:2019der} imply $M_{V,A}\gsim 2$~TeV (for $M_A/M_V>1.1$) and $M_{S_1^1}\gsim 2$~TeV (for $c_d\sim v$).
\end{itemize}
Finally, it is worth to mention some additional conclusions in Ref.~\cite{Pich:2020xzo}, where we have studied how large would be the impact on $\mF_{6,7,9}$ if there were $V$, $A$ and $P$ resonances with masses $M_R\sim 2$~TeV. We found that these LECs could be then as large as $\cO(10^{-3})$--$\cO(10^{-2})$, which would make feasible their detection  
in future experimental analyses.

\end{document}